\documentstyle[aps,epsfig,eqsecnum,prd,preprint]{revtex}
\tightenlines
\begin{document}
\def\non{\nonumber}
\def\la{\langle}
\def\ra{\rangle}
\newcommand{\tld}{\tau(\Lambda_b)/\tau(B^0)}
\newcommand{\tsd}{\tau(B^0_s)/\tau(B^0)}
\newcommand{\hm}{\hat{m}_b}
\newcommand{\rt}{\rho_\tau}
\newcommand{\hr}{\hat{\rho}}
\newcommand{\uv}{/\!\!\!\!\hspace{1pt}v}
\newcommand{\Dslashbot}{{/\!\!\!\!D}\hspace{-3pt}_{\bot}}
\newcommand{\Dbot}{{/\!\!\!\!D}\hspace{-3pt}_{\bot}}
\newcommand{\oDbot}{\overleftarrow{/\!\!\!\!D}\hspace{-5pt}_{\bot}}
\newcommand{\be}{\begin{eqnarray}}
\newcommand{\ee}{\end{eqnarray}}
\newcommand{\vslash}{\not\!{v}}
\newcommand{\eml}{\end{mathletters}}
\newcommand{\bm}{\begin{mathletters}}
\newcommand{\hG}{\hat{\Gamma}^u_0}
\def\Ova{O^q_{V-A}}
\def\Osp{O^q_{S-P}}
\def\Tva{T^q_{V-A}}
\def\Tsp{T^q_{S-P}}

\draft
\title{  $|V_{ub}|$ and $|V_{cb}|$, Charm Counting and Lifetime Differences 
in Inclusive Bottom Hadron Decays}
\author{Y.L Wu\footnote{Email address: ylwu@itp.ac.cn} 
 and Y.A Yan\footnote{Email address: yanya@itp.ac.cn}}
\address{Institute of Theoretical Physics, Chinese Academy of Sciences,
 Beijing 100080, China}
\maketitle
\begin{abstract}
Inclusive bottom hadron decays are analyzed based on the heavy quark effective field 
theory (HQEFT). Special attentions in this paper are paid to the $b\to u$ transitions 
and nonspectator effects. As a consequence, the CKM quark mixing matrix elements 
$|V_{ub}|$ and $|V_{cb}|$ are reliably extracted from the inclusive semileptonic 
decays $B\rightarrow X_{u} e\nu$ and $B\rightarrow X_{c} e\nu$. Various observables, 
such as the semileptonic branch ratio $B_{SL}$, the lifetime differences among 
$B^{-}$, $B^{0}$, $B_s$ and $\Lambda_{b}$ hadrons, the charm counting $n_c$, are 
predicted and found to be consistent with the present experimental data.  
\end{abstract}

\section{Introduction}

 Recently, the heavy quark effective field theory (HQEFT) with keeping both 
 quark and antiquark fields\cite{ylw} have been investigated and applied to 
 both exclusive\cite{wwy1} and inclusive decays\cite{wwy2} of heavy hadrons.
 It have been seen that the contributions and effects from the 
 antiquark field can play a significant role for certain physical observables
 and are also necessary to be considered from the point of view of quantum field theory.
 Consequently, in this new framework of HQEFT, one can arrive at a consistent description on 
 both exclusive and inclusive decays of heavy hadrons. For instance, at zero recoil,
 $1/m_Q$ corrections in both exclusive and inclusive decays are automatically absent 
 when the physical observables are presented in terms of heavy hadron 
 masses\footnote{Note that in the usual heavy quark expansion or 
 in the expansion based on the usual heavy quark effective theory (HQET), 
 $1/m_Q$ corrections in the inclusive decays are absent only when the inclusive decay rate is 
 presented in terms of heavy quark mass ($m_Q$) rather than the heavy hadron mass ($m_H$), 
 the situation seems to be conflict with the case in the exclusive decays where the normalization
 is given in term of heavy hadron mass. Such an inconsistency 
 may be the reason that leads to the difficulty for understanding the lifetime differences 
 among the bottom hadrons.} $m_H$ \cite{wwy1,wwy2}. Our basic point of the considerations is
 based on the fact that a single heavy quark within hadron is off-mass shell 
 by amount of binding energy $\bar{\Lambda}$. Thus a more reliable heavy quark expansion 
 should be carried out in terms of the so-called ``dressed heavy quark" mass 
\[ \hat{m}_Q \equiv 
 \lim_{m_Q \rightarrow \infty} m_H = m_Q + \bar{\Lambda} \]
 with $m_H$ the heavy hadron mass. 
 The new framework of HQEFT developed in \cite{ylw,wwy1,wwy2}  
 enables us to describe a slightly off-mass shell heavy quark within hadrons. 
 Thus such an HQEFT is expected to provide a reliable way to determine the CKM matrix elements 
 $|V_{cb}|$ and $|V_{ub}|$ as well as to explain the lifetime differences of heavy hadrons. 
  
 In this paper we are going to investigate the inclusive bottom hadron decays and will mainly 
pay attention to the analysis on the nonspectator effects and $b \to u $ transitions within 
the framework of HQEFT. One can extract $|V_{cb}|$ either from the end 
point of the differential decay rate of exclusive $B \to D(D^*)$ decays or 
from the total decay rate of inclusive semileptonic decays.
By including nonspectator effects considered in this paper, 
we will present more reliable results for most of the interesting quantities, such as,
$|V_{ub}|$ and $|V_{cb}|$, charm counting $n_c$ and lifetime differences among $B^{-}$, 
$B^{0}$, $B_s$, $\Lambda_b$ hadrons.

   In general, it is not so favorable to extract, in comparison with the extraction  
of $|V_{cb}|$, the CKM matrix element $|V_{ub}|$ due to either experimental or 
theoretical reasons. In experimental side, there exists an overwhelming background 
of $b \to c \ell \bar{\nu}$ decays, its magnitude could be as large as about two orders. 
Even a small leakage of the measurement for $b \to c \ell \bar{\nu}$ decays could affect the 
identification of $b \to u \ell \bar{\nu}$ decays to a large extent. Moreover, 
there still have other background sources such as nonleptonic $b$ decays 
to charmed hadrons which can undergo a semileptonic decay and lead to a leptonic 
misidentification. Thus it may result in a large experimental
error when extracting $|V_{ub}|$ directly from the total decay rate. Another 
method of determining $|V_{ub}|$ is from the lepton 
spectrum at the slice where the energy of the electron $E_e$ is higher than 
$(M_B^2 - M_D^2)/2M_B \approx 2.31~\mbox{GeV}$, since this region can only arise from 
$b \to u \ell \bar{\nu}$ decay. However, only 10\% of the total
identified events of $b \to u e \bar{\nu}$ decays is contained in such a region. Moreover, 
in theoretical side, as such a region lies at the high end point, the bound 
state effect as well as hadronization cannot be neglected in that region. Luckily, 
recent new developments enable one to extract the events of $b\to u$ 
transitions from the dominant $b\to c$ background. The basic point is to 
use the invariant hadronic mass spectrum in the final state, i.e., 
$s_H = (P_b-q)^2$ with $q$ being the momentum of the lepton pair\cite{azm,du}. 
In $b\to u$ semiletonic inclusive decays, more than 90\% events lie in the arrange 
$s_H<m_c^2$. ALEPH collaboration has used this method to identify 
the $b \to u$ events and reported interesting results for 
the $b \to u e \nu_e$ decay\cite{aleph}.

  Another important issue is the nonspectator effects in the bottom hadron decays. 
In the usual heavy quark effective theory (HQET), 
those are the main effects which could result in lifetime differences
among different bottom hadrons $B^0$, $B^0_s$ and $\Lambda_b$. 
Nevertheless, the effects were found to be only less than
$5\%$ of the total decay rates and seem to be too small to explain the experimental data. 

  Our paper is organized as follows: In section 2 we derive the general formalism for the 
total decay widths of $b \to u (c)$ transitions in the framework of HQEFT and also the one 
from nonspectator effects. In section 3 
we provide the numerical results of $|V_{ub}|$ and $|V_{cb}|$ ( $|V_{ub}|/|V_{cb}|$) from 
$b \to u \ell \bar{\nu}$ and $b\to c \ell \bar{\nu}$ semileptonic decays, the charm counting $n_c$ 
with including the nonspectator effects and the results of the ratios 
\be
r_{uu} = {{\cal B}(b \to u \bar{u} d^\prime) \over 
                    {\cal B}(b \to u \ell \bar{\nu})} \,,  \quad 
r_{\tau u} =  {{\cal B}(b \to u \tau \bar{\nu}) \over 
                    {\cal B}(b \to u \ell \bar{\nu})} \,, \quad
r_{cu} = {{\cal B}(b \to u \bar{c} s^\prime) \over 
                    {\cal B}(b \to u \ell \bar{\nu})} \,,
\ee
with $s^\prime = s V_{cs} + d V_{cd}$ and $d^\prime = s V_{us} + d V_{ud}$. We also 
present, to a good approximation, a simplified analytic expression for both $|V_{ub}|$ 
and $|V_{cb}|$ as functions of fundamental parameters $\alpha_s(\mu)$ and $m_c$ as well 
as $\kappa_1$, which may be useful for phenomenological analyses. Our conclusions 
and remarks are presented in the last section.

\section{$b \to u (c) $ Transitions and Nonspectator Effects}

  The decays of bottom hadrons are meditated by the following effective
weak Lagrangian renormalized at the scale $\mu=m_b$
\begin{eqnarray}
   {\cal L}_{\rm eff} &=& - {4 G_F\over\sqrt{2}}\,\sum_{q=u,c}\,V_{qb}\,
    \bigg\{ c_1(m_b)\,\Big[
    \bar d'_L\gamma_\mu u_L\,\bar q_L\gamma^\mu b_L +
    \bar s'_L\gamma_\mu c_L\,\bar q_L\gamma^\mu b_L \Big] \nonumber\\
   &&\phantom{ - {4 G_F\over\sqrt{2}}\,V_{cb}\, }
    \mbox{}+ c_2(m_b)\,\Big[
    \bar q_L\gamma_\mu u_L\,\bar d'_L\gamma^\mu b_L +
    \bar q_L\gamma_\mu c_L\,\bar s'_L\gamma^\mu b_L \Big] \nonumber\\
   &&\phantom{ - {4 G_F\over\sqrt{2}}\,V_{cb}\, }
    \mbox{}+ \sum_{\ell=e,\mu,\tau}
    \bar\ell_L\gamma_\mu\nu_\ell\,\bar q_L\gamma^\mu b_L
    \bigg\} + \mbox{h.c.} \,,
\end{eqnarray}
where $q_L=\frac{1}{2}(1-\gamma_5) q$ denotes a left-handed quark
field, and $d'=d\,V_{ud} + s\,V_{us}$ and $s'=s\,V_{cs} + d\,V_{cd}$.
The Wilson coefficients $c_1$ and $c_2$ at leading-order are
\bm
\begin{eqnarray}\label{cpm}
  & &  c_1 = \frac{1}{2}(c_{+} + c_{-})\ , \qquad c_2 = \frac{1}{2}(c_{+} - c_{-}), \\
  & & c_\pm(m_b) = \left( {\alpha_s(m_W)\over\alpha_s(m_b)}
   \right)^{a_\pm}\  , \qquad   a_- = -2 a_+ = - {12\over 33 - 2 n_f} \,.
\end{eqnarray}
\eml
  Due to optical theorem, the inclusive decay width of bottom hadrons may be expressed 
as the absorptive part of the forward scattering amplitude of bottom hadron $H_b$
\be
\label{ot}
\Gamma(H_b \to X) = {1 \over 2m_{H_b}} ~\mbox{Im} \left(i \int d^4 x<H_b|
 {\cal T}\{ {\cal L}^{(X)}_{eff}(x),{\cal L}^{(X)}_{eff}(0) \}|H_b>\right),
\ee
where ${\cal L}^{(X)}_{eff}(x)$ is the part of the complete $\Delta B = 1$ 
effective Lagrangian which contributes to the particular inclusive final 
state $X$ under consideration. Because the energy released in the process 
is rather large, one may calculate
the inclusive decay width with operator product expansion. For 
non-perturbative part, one can use heavy quark expansion based on the HQEFT, which has 
been shown to be reliable for $b\rightarrow c$ transitions\cite{wwy2} and is
expected to be applicable for $b\rightarrow u$ transitions. 

\subsection{ $b\to u (c)$ Inclusive Decays}

  It is interesting to note that up to the $1/m_b^2$ order
the inclusive decay width of bottom hadrons only depends on two matrix elements 
\bm
\be
A &=& { 1 \over 6m_{H_b}} <H_b|\bar{b} e^{-i m_b \uv v \cdot x} (i D_\bot)^2 
       e^{i m_b \uv v \cdot x} b |H_b>,   \\
N_b &=& {1 \over 12 m_{H_b}}<H_b|\bar{b} g \sigma_{\mu\nu} G^{\mu\nu} b|H_b>, 
\ee
\eml
with $g$ the QCD coupling constant. The decay width for $b \to c$ transitions can be
written in the following general form
\be
\label{general}
\Gamma (H_b \to c + X) = \hat{\Gamma}^c_0 \eta_{cX} \{ I_0(\rho,\rho_X,\hr)+ 
  I_1(\rho,\rho_X,\hr) A + I_2(\rho,\rho_X,\hr) N_b \} ,
\ee
where the functions $\eta_{cX}$ arise from QCD radiative 
corrections. $\hat{\Gamma}^q_0$ is given by
$$\hat{\Gamma}^q_0 \equiv {G_F^2 \hm^5 |V_{qb}|^2 \over 192 \pi^3} ; 
\quad q = c, \ u.$$ with $\hat{m}_b = m_b + \bar{\Lambda}$ the 
``dressed bottom quark" mass. The functions $I_{0}$, $I_{1}$ 
and $I_{2}$ are phase space factors\cite{wwy2} with
$$\rho \equiv m_c^2/\hm^2; \qquad \hr \equiv \hat{m}_c^2/\hm^2\, ,$$ 
here $\hat{m}_c=m_c + \bar{\Lambda}$ is the ``dressed charm quark' mass.
For $b \to c + \ell \bar{\nu}$ semileptonic decays, 
$\rho_X \equiv m^2_\ell/\hm^2$, and
for $b \to c + \bar{u}d^\prime$ decays, $\rho_X \approx 0$ since the light quarks mass 
is much smaller than the bottom quark mass.
For $b\to c + \bar{c} s^\prime$ decays, we take $\rho_X \simeq \hr$ 
which means that the emmitted anti-charm quark $\bar{c}$ is also 
treated as a ``dressed heavy quark". This is slightly different 
from the consideration in ref.\cite{wwy2} where $\rho_X \approx \rho$. 
As a consequence, the charm counting $n_c$ has less dependence on 
the charm quark mass $m_c$. It is noticed that the parameter $\rho$ 
arises from the propagator of 
charm quark and the parameters $\hr$ and $\rho_X$ arise from the 
phase space integral of the differential decay rate $d\Gamma/dy$ 
with $y\equiv 2E_\ell/\hm$ for semileptonic decays and 
$y\equiv 2 E_{\bar{c}(\bar{u})}/\hm$ for nonleptonic decays. 
The integral region of $y$ is
$$2\sqrt{\rho_X} \leq y \leq 1 + \rho_X - \hr.$$

 We now extand the above considerations for $b\to c$ transitions to $b \to u$ 
transitions. To do that, one only needs to notice the differences between  
$b\to c$ and $b\to u$ transitions. One important difference is that 
for $b \to c$ decays the final charm quark $c$ remains to 
be regarded as a ``dressed heavy 
quark". While for $b \to u$ decays, as the masses of $u$ quark and lightest hadrons 
(i.e., pions) are very small, their effects are highly 
suppressed by $1/m_b$ and will be neglected in a good approximation 
$m_u \ll m_\pi \ll \hm$.  
With this consideration, the  differential decay width of $b \to u e \bar{\nu}$ transition 
is simplified 
\begin{eqnarray}
{1 \over \hG}{d\Gamma \over dy} = 2(3-2y)y^2 - {6N_b \over \hm^2}
   \{3y^2-\delta(y-1)\} + {2A \over \hm^2}\{3\,y^2+ 2\delta(y-1)
   +2\delta^\prime(y-1) \}
\end{eqnarray}
with $y\equiv 2E_e/\hm$.
The integral region of the phase space is
$$0 \leq y \leq 1-m_\pi^2/\hm^2 + \rho_e,$$
which indicates that the $\delta$-functions cannot contribute to the total decay width as $y<1$. 
The $b \to u e \bar{\nu}$ decay width is simply given by
\begin{eqnarray}
{\Gamma \over \hG} = 1 + {2 A \over \hm^2} - {6 N_b \over \hm^2} 
  + O({1 \over \hm^3}),
\end{eqnarray}
where we have neglected the terms of $O(m_\pi^2/\hm^2)$.

  When including the one-loop QCD corrections, we have the following 
general forms for $b \to u$ transitions 
\bm
\label{butran}
\begin{eqnarray}
\label{busl}
{\Gamma(b \to u \ell \bar{\nu}) \over \hG} &&= 
  \{ 1- {2 \over 3\pi}(\pi^2 - {25 \over 4}) \alpha_s(\mu)\} 
  (1 + {2 A \over \hm^2} - {6 N_b \over \hm^2});     \quad \ell = e, \ \  \mu,           \\
{\Gamma(b \to u \tau \bar{\nu}) \over \hG} && = \{1- 0.665 \alpha_s(\mu)\}
         (f(\rho_\tau)+ {2A \over \hm^2} (1-\rt)^4 - {6 N_b \over \hm^2}(1 + \rt)^3),  \\
{\Gamma(b \to u \bar{u} d^\prime) \over \hG} &&= \{1-\eta_{uu}\alpha_s(\mu)\}
   \{1 + {2A \over \hm^2} - {6 N_b \over \hm^2} \}+ (c_+^2(\mu) - c_-^2(\mu)) {6N_b \over \hm^2},
    \\
{\Gamma(b \to u \bar{c} s^\prime) \over \hG} &&= \{1-\eta_{uc} \alpha_s(\mu)\}
   \{f(\hr)+ {2A \over \hm^2} (1-\hr)^4 - {6 N_b \over \hm^2}(1 - \hr)^3 \} 
   \nonumber \\
   && + (c_+^2(\mu) - c_-^2(\mu))
     {6 ( 1 - \hr)^3 N_b \over \hm^2} ,
\end{eqnarray}
\end{mathletters}
with 
$$\rt \equiv m^2_\tau/\hm^2; \quad \hr \equiv \hat{m}_c^2/\hat{m}^2_b ;
  \quad    d^\prime = d V_{ud} + s V_{us} ,$$
and 
$$f(x) = 1 - 8x + 8x^3 - x^4 -12 x^2 \ln x .$$

The two loop QCD corrections for $b\to u \ell \nu$ decays have 
recently been carried out in ref.\cite{timo}. The one-loop QCD 
corrections for $b\to u \bar{u} d$ decays can be obtained from 
the ones for $b \to c$ transitions given in 
refs.\cite{Bagan,MBV,AC} by simply taking the limit $a \equiv 
m_c^2/m_b^2 \to 0$. And the one loop QCD corrections for 
$b \to u\bar{c} s$ decays have been calculated in ref.\cite{jam}. 
For consistent, we shall use the one loop results for 
$b \to u \ell \nu$ decays to calculate the ratios $r_{uu}$, $r_{\tau u}$ 
and $r_{cu}$, and adopt the two loop QCD corrections for the semileptonic 
decays $b \to q \ell \nu\, (q=c\,,u)$ to extract the CKM matrix elements
$|V_{ub}|$ and $|V_{cb}|$ as well as the ratio $|V_{ub}|/|V_{cb}|$. 

\subsection{Nonspectator Effects in Bottom Hadron Decays}

To order of $1/m_b^3$, the nonspectator effects due to Pauli 
interference and
$W$-exchange ,
$B_i$, $\varepsilon_i$ ($i=1, 2$), $\tilde B$, and $r$, may have sizeable contributions to the
lifetime differences of bottom hadrons due to a phase-space enhancement by a factor of $16\pi^2$.
The four-quark operators relevant to inclusive nonleptonic $B$ decays are
\bm
\begin{eqnarray}\label{4qops}
   O_{V-A}^q &=& \bar b_L\gamma_\mu q_L\,\bar q_L\gamma^\mu b_L
    \,, \\
   O_{S-P}^q &=& \bar b_R\,q_L\,\bar q_L\,b_R \,, \\
   T_{V-A}^q &=& \bar b_L\gamma_\mu t^a q_L\,
\bar q_L\gamma^\mu  t^a b_L \,, \\
   T_{S-P}^q &=& \bar b_R\,t^a q_L\,\bar q_L\, t^ab_R \,,
\end{eqnarray}
\eml
where
$q_{R,L}={1\pm\gamma_5\over 2}q$ and $t^a=\lambda^a/2$ with $\lambda^a$
being the Gell-Mann matrices. For the matrix elements of these 
four-quark operators between $B$ hadron states,
we follow the definitions given in ref.\cite{NS}
\bm
\be \label{parameters}
{1\over 2m_{ B_q}}\la \bar B_q|\Ova|\bar B_q\ra &&\equiv {f^2_{B_q} m_{B_q}
\over 8}B_1\,, \\
{1\over 2m_{B_q}}\la \bar B_q|\Osp|\bar B_q\ra &&\equiv {f^2_{B_q} 
m_{B_q}\over 8}B_2\,,\\
{1\over 2m_{B_q}}\la \bar B_q|\Tva|\bar B_q\ra &&\equiv {f^2_{B_q} 
m_{B_q}\over 8}\varepsilon_1\,, \\
{1\over 2m_{B_q}}\la \bar B_q|\Tsp|\bar B_q\ra &&\equiv {f^2_{B_q} 
m_{B_q}\over 8}\varepsilon_2\,,\\
{1\over 2m_{\Lambda_b}}\la \Lambda_b |\Ova|\Lambda_b \ra 
&&\equiv -{f^2_{B_q} m_{B_q}\over 48}r\,,\\
{1\over 2m_{\Lambda_b}}\la \Lambda_b |\Tva|\Lambda_b \ra
&&\equiv -{1\over 2} (\tilde B+{1\over 3})
{1\over 2m_{\Lambda_b}}\la \Lambda_b |\Ova|\Lambda_b \ra \,,
\ee
\eml
where $f_{B_q}$ is the $B_q$ meson decay constant defined via
\begin{equation}
\langle 0|\bar q\gamma_\mu \gamma_5 b|\bar B_q(p)\rangle =i\,f_{B_q} p_\mu \,.
\end{equation}
Under the factorization approximation, $B_i=1$ and $\varepsilon_i=0$, and
under the valence quark approximation $\tilde B=1$.

  Applying the treatment in ref.\cite{NS}, the decay widths due to nonspectator effects 
have the following form in our present considerations 
\bm
\be
\label{mesons}
   {1\over 2 m_B}\,\langle B^-|\,{\bf\Gamma}_{\rm spec}\,
   |B^-\rangle &&= \hG\,\hat{\eta}_{\rm spec}\,(1-\hr)^2\,\Big\{
    (2 c_+^2 - c_-^2)\,B_1 + 3 (c_+^2 + c_-^2)\,\varepsilon_1 \Big\}
    \,,  \\
   {1\over 2 m_B}\,\langle B_d|\,{\bf\Gamma}_{\rm spec}\,|B_d\rangle
    &&= - \hG\,\hat{\eta}_{\rm spec}\,(1-\hr)^2\,|V_{ud}|^2 \,
    \Bigg\{ {1\over 3}\,(2 c_+ - c_-)^2\,\bigg[
    \bigg( 1 + {\hr\over 2} \bigg)\,B_1 - (1+2\hr)\,B_2 \bigg]
    \nonumber \\
   &&\quad\mbox{}+ {1\over 2}\,(c_+ + c_-)^2\,\bigg[
    \bigg( 1 + {\hr\over 2} \bigg)\,\varepsilon_1
    - (1+2\hr)\,\varepsilon_2 \bigg] \Bigg\} \nonumber \\
   &&\quad\mbox{}- \hG\,\hat{\eta}_{\rm spec}\,\sqrt{1-4\hr}\,
    |V_{us}|^2\,\Bigg\{ {1\over 3}\,(2 c_+ - c_-)^2\,\Big[
    (1-\hr)\,B_1 - (1+2\hr)\,B_2 \Big] \nonumber \\
   &&\quad\mbox{}+ {1\over 2}\,(c_+ + c_-)^2\,\Big[
    (1-\hr)\,\varepsilon_1 - (1+2\hr)\,\varepsilon_2 \Big] \Bigg\} \,, \\
\label{baryon}
   {1\over 2 m_{\Lambda_b}}\,\langle\Lambda_b|\,
   {\bf\Gamma}_{\rm spec}\,|\Lambda_b\rangle
   &&= \hG\,\hat{\eta}_{\rm spec}\,{r\over 16}\,\Bigg\{
    4 (1-\hr)^2\,\Big[ (c_-^2 - c_+^2)
    + (c_-^2 + c_+^2)\,\widetilde B \Big]\nonumber  \\
   &&\mbox{}- \Big[ (1-\hr)^2 (1+\hr)\,|V_{ud}|^2
    + \sqrt{1-4 \hr}\,|V_{us}|^2 \Big] \nonumber \\
   &&\quad\times \Big[ (c_- - c_+)(5 c_+ - c_-)
    + (c_- + c_+)^2\,\widetilde B \Big] \Bigg\} \,.
\ee
\eml
where $c_\pm=c_1\pm c_2$, and
\begin{equation}\label{Gamma0}
  \hat{\eta}_{\rm spec} = 16\pi^2\,{f_B^2 m_B\over \hm^3} \,. \nonumber
\end{equation}
The spectator contribution to the width of $B_s$ meson is simply
obtained from that of the $B_d$ meson by the replacement:
$|V_{ud}| \leftrightarrow |V_{us}|$, and $(f_B,m_B)\to
(f_{B_s},m_{B_s})$. Strictly speaking, the values of the parameters $B_i$ and
$\varepsilon_i$ for the $B_s$ meson should be different from those for
the $B_d$ meson due to SU(3)-breaking effects.

\section{Numerical Analysis}

\subsection{Basic Formulae }

 It has been shown from above section that the leading order nonperturbative corrections
for $b \to u$ transitions only involve two matrix elements  at the order of
$1/m^2_b$. In the framework of new formulation of HQEFT, the mass formulae for the 
hadrons containing a single heavy quark are
\be
m_H &=& m_Q + \bar{\Lambda} - {<H_v| \bar{Q}^{(+)}_v (i \ \Dslashbot)^2 
      Q^{(+)}_v |H_v> \over 2\bar{\Lambda} \cdot m_Q}  + O({1 \over m^2_Q}).
\ee
Define 
\bm
\be
\label{lambda}
\kappa_1 & \equiv & -<H_v|\bar{Q}^{(+)}_v D_\bot^2 
	Q^{(+)}_v|H_v>/(2\bar{\Lambda})  ,\\
\kappa_2 & \equiv & - <H_v|\bar{Q}^{(+)}_v g \sigma_{\mu\nu} 
G^{\mu\nu} Q^{(+)}_v|H_v>/(4d_H \bar{\Lambda}).
\ee
\eml
with $d_H = -3$ for pseudoscalar mesons, $d_H = 1$ for vector mesons and
$d_H = 0$ for ground state heavy baryons, the mass formulae can
be reexpressed as
\be
\label{mass}
m_H = m_Q + \bar{\Lambda} - {\kappa_1 \over m_Q} + {d_H \kappa_2 \over  m_Q}
   + O({1 \over m^2_Q}).
\ee

 Thus the ``dressed bottom quark" mass $\hm$ can be 
rewritten in terms of the hadron mass 
\be
\label{addmass}
\hat{m}_b &=& m_{H_b}+{\kappa_1-d_H \kappa_2 \over m_b} +O({1 \over m^2_b})   
   \nonumber   \\
&=& m_{H_b}+{\kappa_1-d_H \kappa_2 \over m_{H_b}} +O({1 \over m^2_{H_b}}).
\ee

  Using eq.(\ref{mass}), the value of $\kappa_2$ can be directly extracted 
from the known $B - B^*$ mass splitting 
\be
\kappa_2 \simeq {1 \over 8} (m_{B^{*0}}^2 - m_{B^0}^2) = 0.06~\mbox{GeV}^2,
\ee
which has an accuracy up to the power correction of $\bar{\Lambda}/2m_b \sim 5\%$. 
The value of $\kappa_1$ depends on bottom quark mass and binding energy, only a reliable 
range of $\kappa_1$ can be obtained. In the following analysis we shall take the range

$$ -0.6~\mbox{GeV}^2 \leq \kappa_1 \leq -0.1~\mbox{GeV}^2.$$ 

  For the two matrix elements $A$ and $N_b$, we only need to consider the 
leading order terms in $1/m_b$ when the decay rates are evaluated up to $1/m^2_b$ order.
It is then not difficult to yield
\be
A = {\kappa_1 \over 3} \,, \quad
N_b = {d_H \kappa_2 \over 3}\,.
\ee

  The magnitudes of nonspectator effects depend on the values of $\varepsilon_i$ and $B_i$. 
From theoretical calculations\cite{NS,H.Y,ne1,ne2,pf}, there remain large uncertainties.
For a conservative consideration, we may take following values for those parameters
\begin{eqnarray}
\label{para}
B_1 (m_b) & \simeq & B_2 (m_b) \simeq 1 \,, \nonumber \\
\varepsilon_1 (m_b) & \simeq & \varepsilon_2 (m_b) = -0.10\pm 0.05\,,  \\
r & \simeq & 0.3\,, \quad \tilde{B} \approx 1\,. \nonumber
\end{eqnarray}

\subsection{$|V_{ub}|$, $|V_{cb}|$ and $|V_{ub}|/|V_{cb}|$ from 
    Inclusive Semileptonic Decays}

 With the above analyses, the two important CKM matrix elements $|V_{ub}|$ and $|V_{cb}|$ can be 
extracted from inclusive semileptonic decays. It is seen that up to $1/m^2_b$ order 
the $b \to c$ and $b \to u$ semileptonic decays mainly relate to the variables $\kappa_1$ and
$\kappa_2$ as well as the energy scale $\mu$ and charm quark mass $m_c$. Fixing $\kappa_2 =0.06$ and
fitting  to the current experimental data for total decay rates of the inclusive $b \to c$ and $b \to u$ 
semileptonic decays, we find that $|V_{ub}|$ and  $|V_{cb}|$ may be given, to a good approximation, 
by the following form
\bm
\label{dert}
\begin{eqnarray}
\label{Vub}
|V_{ub}| =&& 3.45 \times 10^{-3}\, \{ 1 - 
  0.02 (1 - {\kappa_1 \over -0.2~\mbox{GeV}^2}) +
   4 \times 10^{-4} (1 - {\kappa_1 \over -0.2~\mbox{GeV}^2})^2\}  \nonumber \\
 \times &&
    [1 - 2.40 (\alpha_s - 0.3) - 2.04 (\alpha_s - 0.3)^2]^{-{1 \over 2}}
\left( {{\cal B}(B^0 \to X_u e \bar{\nu}) \over 0.00173}\right)^{1 \over 2} 
   \left({1.56~\mbox{ps} \over \tau(B^0) } \right)^{1 \over 2} , \\
|V_{cb}| =&& 3.93 \times 10^{-2}\, \{ 1 - 
     4\times 10^{-3} (1 - {\kappa_1 \over -0.2~\mbox{GeV}^2})^2  - 
    (1 - {\kappa_1 \over -0.2~\mbox{GeV}^2}) \nonumber \\
\times &&    [0.038 + 0.091 (1 - {m_c \over 1.75~\mbox{GeV}}) ] 
     - 0.7 (1 - {m_c \over 1.75~\mbox{GeV}}) - 
    0.18 (1 - {m_c \over 1.75~\mbox{GeV}})^2 \} \nonumber \\
\times &&   [1 - 2.05 (\alpha_s - 0.3)
   - 2.17 (\alpha_s - 0.3)^2]^{-{1 \over 2}}
\left( {{\cal B}(B^0 \to X_c e \bar{\nu}) \over 0.1048}\right)^{1 \over 2} 
   \left({1.56~\mbox{ps} \over \tau(B^0) } \right)^{1 \over 2} .
\end{eqnarray}
\eml

  Let us now discuss possible theoretical uncertainties for the quantities 
$|V_{ub}|$ and $|V_{cb}|$ or $|V_{ub}|/|V_{cb}|$. 

  (1). From discarding high order nonperturbative corrections. In the present
analysis we have only expanded the matrix element to $1/m_b^2$ order. It has been noted that 
there is no $1/m_b$ order corrections and the magnitude of $1/m_b^2$ corrections is less
than 5.5\%. Since the higher order nonperturbative corrections are expected  
to be much smaller than the ones at $1/m_b^2$ order and their effects appear to be no more than 1\% in 
a conservative estimation. 

  (2). From the hadronic matrix elements, i.e., $\kappa_1$, $\kappa_2$ and 
the charm quark mass $m_c$. As mentioned above, the extraction of $\kappa_2$ could have an 
accuracy up to 5\%, while the value of $\kappa_1$ has not been well determined. 
As the ``dressed bottom quark" mass entered to the decay rates in powers of $\hat{m}_Q^5$, 
the uncertanities of $\kappa_1$ become the main sources of the uncertainty. 
From Fig. 1 and 
Fig. 2 one can see that the resulting uncertainties are no more than 3\% for 
$|V_{ub}|$ and 2.7\% for $|V_{ub}|/|V_{cb}|$. As for the charm quark mass $m_c$, 
we have considered the range $1.55~\mbox{GeV} \leq m_c \leq 1.80~\mbox{GeV}$, it leads 
to an uncertainty of 3.5\% for $|V_{ub}|/|V_{cb}|$  (here we have limited the case  
that $m_c \leq 1.65~\mbox{GeV}$ for $\kappa_1 = -0.6~\mbox{GeV}^2$ and 
$m_c \leq 1.8~\mbox{GeV}$ for $\kappa_1 = -0.5~\mbox{GeV}^2$ from the consideration of 
lifetime differences among $B^0$ and $B^0_s$ mesons as well as $\Lambda_b$ baryon). 

  (3). From the perturbative QCD corrections. The first order QCD corrections have been 
calculated in ref.\cite{AC} and the second order results have been presented in 
ref.\cite{MM,LLSS,9903226}. Since the size of the second order QCD corrections 
have been found to be comparable with the first order ones, the unknown higher order 
QCD corrections may lead to a sizable theoretical uncertainties. 
The additional uncertainties could arise from the energy scale $\mu$. 
Generally, the scale $\mu$ in the $b$ decays is taken from 
$m_b/2$ to $2m_b$, which could lead to a large theoretical uncertainty for the 
determination of $|V_{ub}|$. One method to reduce the possible uncertainties
arising from the perturbative corrections is to choose a proper value of $\mu$
by fitting the experimental data of the semileptonic decays 
${\cal B}(b \to c \ell \bar{\nu})$. When only one-loop QCD 
corrections are considered, it is seen that one should take lower values of $\mu$ with the 
range $m_b/4 \leq \mu \leq m_b$. From Fig. 1 one can see that this would lead to an 
uncertainty of about 3\% for $|V_{ub}|$ and a similar one for $|V_{ub}|/|V_{cb}|$.

  With above considerations and using the ALEPH experimental data\cite{aleph} without 
the non-Gaussian errors, we obtain the following results
\bm
\label{numv}
\be
&&|V_{ub}| = (3.48  \pm 0.11_{th} \pm 0.62_{exp}) \times 10^{-3}, \\
&&|V_{cb}| = (3.89 \pm 0.20_{th} \pm 0.05_{exp}) \times 10^{-2}, \\
&&|V_{ub}|/|V_{cb}| = 0.089 \pm 0.005_{th} \pm 0.015_{exp}.
\ee
\eml
by choosing the parameters $m_c$ and $\kappa_1$ to be within the range:
$1.55~\mbox{GeV} \leq m_c \leq 1.80~\mbox{GeV}$, 
$-0.6~\mbox{GeV}^2 \leq \kappa_1 \leq -0.1~\mbox{GeV}^2$ and considering the lifetime
ratio $\tld$ to be $0.70 \leq \tld \leq 0.85$ as well as the ratio between the $\tau$
and $\beta$ decay $B_r(b\to c \tau \nu)/Br(b\to c e \nu) \leq 0.285$.

\subsection{Ratios $r_{uu}$, $r_{\tau u}$ and $r_{c u}$ 
}

  The nonperturbative corrections in the $b \to u$ transitions have a very simple form and their 
effects are also quite small, thus the uncertainties induced by them are much smaller 
than those of the perturbative corrections. Fixing  
${\cal B}(b \to c e \bar{\nu})$ to be 10.48\%, the ratios defined in eq. (1.1) 
are found to be
\bm
\be
r_{uu}     &=& 4.8  \pm 0.5 \  , \\
r_{cu}     &=& 2.4  \pm 0.3  \  , \\
r_{\tau u} &=& 0.44 \pm 0.02 \  .
\ee
\eml
In obtaining the above results, the range 
$ -0.6 \mbox{GeV}^2 \leq \kappa_1 \leq -0.1~\mbox{GeV}^2$ has been used. 
The uncertainties mainly arise from that of the energy 
scale $\mu$. It is interesting to note that  the ratio between $\tau$ and 
$\beta$ decays approaches to be about half in the $b \to u$ transitions, 
while it is only about quarter in the $b \to c$ transitions.

\subsection {Nonspectator Effects}

   The contributions from the nonspectator effects could vary in the decays of 
different bottom hadrons. Using the values given in eq.(\ref{para}) for various parameters, 
the magnitude could be from $-6\%$ to $-11\%$ in $B^+$ decays, and about (0.6-0.7)\%, 
(0.2-0.3)\% and (0.6-0.75)\% in $B^0$, $B_s^0$ and $\Lambda_b$ decays, respectively. 

  As a consequence, we arrive at the following results for the ratios of the lifetimes
\bm
\be
&& \frac{\tau (B^-)}{\tau(B_d)} = 1.08 \pm 0.05\,,   \\
\label{tsd}
&& \frac{\tau (B_s)}{\tau(B_d)} = 0.96 \pm 0.06\,,  \\
\label{tld}
&& \frac{\tau (\Lambda_b)}{\tau (B_d)} = 0.78 \pm 0.05 \,.
\ee
\eml
which are in good agreement with the experimental data.

 Note that the contributions to the ratios in eqs.(\ref{tsd}) and (\ref{tld}) from 
the nonspectator effects are rather small when the relevant parameters take values 
in eq.(\ref{para}). Thus the usual HQET fails in explaining the 
lifetime differences among $B^0$, $B_s^0$ and $\Lambda_b$ hadrons. This is because 
the lifetime differences in the usual HQET arise mainly from the nonspectator effects.

\subsection{More Numerical Results}

  We show in Fig.3 the correlation between the charm counting $n_c$ and the branching ratio $B_{SL}$ 
of the semileptonic $B\rightarrow X_c e\nu$ decay with different values of the charm quark mass 
$m_{c}$ ($m_{c}=1.55 \sim 1.80$ GeV), the energy scale $\mu$ ($\mu =m_{b}/2 \sim 2m_{b}$) and 
the parameter $\kappa_1$ (the solid curve for $\kappa_1 = -0.5 GeV^2$ and the dotted curve for 
$\kappa_1 = -0.2 GeV^2$). It is seen that the would average values of the charm counting $n_{c}$ and 
the branch ratio $B_{SL}$ lie in the allowed region predicted from the HQEFT. The lower value of 
$\mu=m_b/2 \sim m_b$ and larger value of $m_c = 1.65\sim 1.80$ GeV as well as the smaller value of
$|\kappa_1|$ seem to be favorable. In Fig.4, we present the correlation among the three 
observables: the charm counting $n_c$, the branching ratio $B_{SL}$ and the lifetime ratio $\tld$. 
It is interesting to notice that the stable region is more favorable to the experimental data. 

  It is also seen from Fig. 3 that the charming counting $n_c$ has strong dependance on charm quark 
mass $m_c$ and $\kappa_1$. Within the allowed range of $m_c$ and $\kappa_1$, and by considering the 
lifetime ratio $\tld$ to be $0.70 \leq \tld \leq 0.85$ as well as the ratio between the $\tau$
and $\beta$ decay to be $B_r(b\to c \tau \nu)/Br(b\to c e \nu) \leq 0.285$, the charming $n_c$ 
is found to be
\be
n_c = 1.19 \pm 0.04\,.
\ee
It is notice that small values of $\kappa_1$ and large $m_c$ will result in a low value of 
charm counting $n_c = 1.15$.

 For more clear, we provide in Table 1 and Table 2 the most reliable values for the various interesting 
 observables. The agreement with the experimental data must be regarded as a success of QCD 
 since both the perturbative corrections and nonperturbative contributions described by HQEFT 
 are resulted from QCD.  We hope that a better agreement can be arrived by considering 
 higher order contributions.

{\bf Table 1.} The quantities $V_{ub}$, $V_{cb}$ and lifetime ratios among bottom hadrons as well as 
the relative contributions between $b\rightarrow u + X_i$ (with $X_i = ud', cs', 
\tau \nu$) and $b\rightarrow u + e\nu$ transitions are given as functions of $m_c$ and $\kappa_1$. 
For the given values of $m_c$ and $\kappa_1$, the value of $\mu$ is yielded
by fixing the semileptonic branching ratio $B_{SL} = 10.48\% $. 
The quantities except $r_{\tau u}$, $r_{cu}$ and $r_{uu}$ have been 
evaluated by including two-loop QCD corrections and nonspectator effects.

\begin{center}
\begin{tabular}{|c|c|c|c|c|c|c|c|c|c|c|c|c|} \hline
$m_c$(~\mbox{GeV})   & \multicolumn{3}{|c|}{ 1.55 } 
   & \multicolumn{3}{|c|}{ 1.65 }
   & \multicolumn{3}{|c|}{ 1.75 }
   & \multicolumn{3}{|c|}{ 1.80 }   \\  \cline{1-13}
$\kappa_1(~\mbox{GeV}^2)$   
   & $-0.2$   & $-0.4$ & $-0.6$  &
     $-0.2$   & $-0.4$ & $-0.6$ &
     $-0.2$   & $-0.4$ & $-0.6$ &
     $-0.2$   & $-0.4$ & $-0.6$  \\ \hline
$\mu(~\mbox{GeV})$ & 
  2.36 & 2.53 & 2.73 & 2.27 & 2.43 & 2.69 & 2.19 & 2.36 & 
  2.74 & 2.14 & 2.34 & 2.82     \\ \hline
$|V_{ub}|(10^{-3})$
  & 3.42 & 3.47 & 3.52 & 3.43 & 3.48  & 3.52 & 3.44 & 3.49  & 3.52 
  & 3.45  & 3.49 & 3.51  \\ \hline
$|V_{cb}|(10^{-2})$
  & 3.59 & 3.71 & 3.83 & 3.75 & 3.87  & 3.88 & 3.92 & 3.99  & 3.81 
  & 4.01  & 4.03 & 3.72  \\ \hline
$|V_{ub}|/|V_{cb}|(10^{-2})$
  & 9.51 & 9.34 & 9.19 & 9.15 & 9.00  & 9.08 & 8.77 & 8.74  & 9.23 
  & 8.59  & 8.66 & 9.43  \\ \hline
${\tau(B^0_s) \over \tau(B^0)}$
  & 0.92 & 0.92 & 0.93  & 0.92& 0.92 & 0.98  & 0.92 & 0.94  
  & 1.06 & 0.92  & 0.96 & 1.12  \\ \hline
${\tau(\Lambda_b) \over \tau(B^0)}$
  & 0.76 & 0.74 & 0.73 & 0.75  & 0.73 & 0.78  & 0.73 & 0.75 
   & 0.88 & 0.72  & 0.76 & 0.97  \\ \hline
${\tau(B^+) \over \tau(B^0)}$
  & 1.07 & 1.07 & 1.07 & 1.08  & 1.08 & 1.07  & 1.09 & 1.09 
   & 1.06 & 1.10  & 1.09 & 1.05  \\ \hline
$n_c$
  & 1.19 & 1.21 & 1.23 & 1.18  & 1.20 & 1.22  & 1.17 & 1.19 
   & 1.23 & 1.17  & 1.19 & 1.23  \\ \hline
$r_{uu}$
  & 4.94 & 4.79 & 4.64 & 5.00 & 4.85  & 4.66 & 5.06 & 4.89  & 4.63 
  & 5.09  & 4.91 & 4.60  \\ \hline
$r_{cu}$
  & 2.26 & 2.36 & 2.45 & 2.30 & 2.41  & 2.47 & 2.36 & 2.45  & 2.45 
  & 2.38  & 2.46 & 2.42  \\ \hline
$r_{\tau u}$
  & 0.46 & 0.45 & 0.45 & 0.46 & 0.45  & 0.45 & 0.46 & 0.45  & 0.45 
  & 0.46  & 0.45 & 0.45  \\ \hline
\end{tabular}
\end{center}

\begin{center}
Table 2. The quantities $V_{ub}$, $V_{cb}$ and lifetime ratios among bottom hadrons as well as 
the relative contributions between $b\rightarrow u + X_i$ (with $X_i = ud', cs', 
\tau \nu$) and $b\rightarrow u + e\nu$ transitions are given as functions of $m_c$ and $\kappa_1$. 
For the given values of $m_c$ and $\kappa_1$, the value of $\mu$ is fixed to be $\mu=2.5~\mbox{GeV}$. 
The quantities except $r_{\tau u}$, $r_{cu}$ and $r_{uu}$ have been 
evaluated by including two-loop QCD corrections and nonspectator effects.
\end{center}
\begin{center}
\begin{tabular}{|c|c|c|c|c|c|c|c|c|c|c|c|c|} \hline
$m_c$(~\mbox{GeV})   & \multicolumn{3}{|c|}{ 1.55 } 
   & \multicolumn{3}{|c|}{ 1.65 }
   & \multicolumn{3}{|c|}{ 1.75 }
   & \multicolumn{3}{|c|}{ 1.80 }   \\  \cline{1-13}
$\kappa_1(~\mbox{GeV}^2)$   
   & $-0.2$   & $-0.4$ & $-0.6$  &
     $-0.2$   & $-0.4$ & $-0.6$ &
     $-0.2$   & $-0.4$ & $-0.6$ &
     $-0.2$   & $-0.4$ & $-0.6$  \\ \hline
$B_{SL}(\%)$
  & 10.70 & 10.44 & 10.18 & 10.85 & 10.59  & 10.23 & 11.01 & 10.70  & 10.16 
  & 11.09  & 10.73 & 10.08  \\ \hline
$|V_{ub}|(10^{-3})$
  & 3.40 & 3.47 & 3.54 & 3.40 & 3.47  & 3.54 & 3.40 & 3.47  & 3.54 
  & 3.40  & 3.47 & 3.54  \\ \hline
$|V_{cb}|(10^{-2})$
  & 3.57 & 3.72 & 3.86 & 3.71 & 3.85  & 3.91 & 3.86 & 3.96  & 3.85 
  & 3.94  & 4.00 & 3.77  \\ \hline
$|V_{ub}|/|V_{cb}|(10^{-2})$
  & 9.53 & 9.33 & 9.17 & 9.17 & 9.01  & 9.06 & 8.81 & 8.76  & 9.21 
  & 8.64  & 8.68 & 9.41  \\ \hline
${\tau(B^0_s) \over \tau(B^0)}$
  & 0.92 & 0.92 & 0.93  & 0.92& 0.92 & 0.98  & 0.92 & 0.94  
  & 1.07 & 0.92  & 0.96 & 1.13  \\ \hline
${\tau(\Lambda_b) \over \tau(B^0)}$
  & 0.76 & 0.74 & 0.73 & 0.74  & 0.73 & 0.78  & 0.73 & 0.74 
   & 0.88 & 0.72  & 0.76 & 0.97  \\ \hline
${\tau(B^+) \over \tau(B^0)}$
  & 1.07 & 1.07 & 1.07 & 1.07  & 1.08 & 1.07  & 1.08 & 1.08 
   & 1.06 & 1.08  & 1.08 & 1.05  \\ \hline
$n_c$
  & 1.19 & 1.21 & 1.23 & 1.18  & 1.20 & 1.23  & 1.17 & 1.19 
   & 1.23 & 1.16  & 1.19 & 1.24  \\ \hline
$r_{uu}$
  & 4.85 & 4.81 & 4.76 & 4.85 & 4.81  & 4.76 & 4.85 & 4.81  & 4.76 
  & 4.85  & 4.81 & 4.76  \\ \hline
$r_{cu}$
  & 2.19 & 2.37 & 2.56 & 2.19 & 2.37  & 2.56 & 2.19 & 2.37  & 2.56 
  & 2.19  & 2.37 & 2.56  \\ \hline
$r_{\tau u}$
  & 0.46 & 0.45 & 0.45 & 0.46 & 0.45  & 0.45 & 0.46 & 0.45  & 0.45 
  & 0.46  & 0.45 & 0.45  \\ \hline
\end{tabular}
\end{center}

\section{Conclusions and Remarks}

  Based on the new framework of HQEFT with including antiquark 
  contributions\cite{ylw,wwy1,wwy2}, we have made a systematic analysis for 
  the $b\rightarrow u (c)$ transitions with including the nonspectator effects.
  The important CKM quark mixing matrix elements $|V_{ub}|$ and $|V_{cb}|$ have been 
  reliably extracted from the inclusive semileptonic decays $B\rightarrow X_{u} e\nu$ and 
  $B\rightarrow X_{c} e\nu$. The resulting predictions for the various observables, 
  such as the semileptonic branch ratio $B_{SL}$, the lifetime differences among $B^{-}$, $B^{0}$, 
  $B_s$ and $\Lambda_{b}$ decays, and the charm counting $n_c$ are consistent with 
  the present experimental data within the allowed region of parameters.   
     
 We would like to remark that it was thought before that the results like eqs.(\ref{dert}) 
 may not be suitable to extract the CKM matrix element $|V_{ub}|$ and the ratio $|V_{ub}|/V_{cb}|$ 
due to the difficulty of identifying the events of $b \to u \ell \nu$ from 
the overwhelming $b \to c \ell \nu$ background. However, the situation has been changed because of 
the progresses of the technique of using the invariant hadronic mass spectrum\cite{azm,du,aleph}.
Especially, the ALEPH collaboration\cite{aleph} has made it feasible. 
Our results given in eqs.(\ref{dert}) and (\ref{numv}) have been
obtained by using the ALEPH data. Such results are more close to those obtained by using the 
parton model\cite{jin}, while they are somewhat smaller than 
those predicted from the usual HQET. It has been seen that the $b \to u$ transitions have 
different features in comparison with the $b \to c$ transitions, for instance, the ratio between 
the $\tau$  and $\beta$ decays in the $b \to u$ transitions is larger by a factor of two
than the one in the $b \to c$ transitions. The nonspectator effects are in general not 
large enough to understand the observed lifetime ratio $\tld$ in the usual HQET, except one chooses 
unexpected large values of $r$ and $\tilde{B}$, for such a choice, the nonspectator effects 
in $B^0$ decays are still small and their contributions to the total decay 
width are generally less than 1\%, but the nonspectator effects in the $\Lambda_b$ decays must become 
unexpected large and their contributions to the total decay width have to be
about 20\% at the order of $1/m_b^3 $ in order to explain the observed lifetime ratio 
$\tld$ in the usual HQET. As a consequence, the heavy quark expansion in the usual HQET 
may become unreliable if one insists such an explanation for the lifetime difference between 
$B^{0}$ and $\Lambda_b$. 

 It is very interesting to further explore the applications of the HQEFT with keeping both 
 quark and antiquark fields\cite{ylw}, in particular for the processes concerning the 
 quark and antiquark annihilations and productions, a special case with 
 kinematic regimes of heavy quark pair production near the threshold 
 has recently been discussed in ref.\cite{BS}.

\acknowledgments

We would like to thank Prof. Y. B. Dai and Dr. C. Liu for useful discussions. 
This work was supported in part by the NSF of China under the grant No. 19625514.



\begin{figure}
\label{fig1}
\caption{$|V_{ub}|$ as function of $\mu$ and $\kappa_1$.}
\end{figure}

\begin{figure}
\label{fig2}
\caption{$|V_{ub}|/|V_{cb}|$ as function of $m_c$ and $\kappa_1$. In this figure 
we have chosen $\mu$ to normalize $B_r(B^0 \to c \ell \nu) = 10.48\%$. }
\end{figure}

\begin{figure}
\label{fig3}
\caption{The correlation between $B_{SL}$ and $n_c$. The solid line is for 
$\kappa_1 = -0.5 ~\mbox{GeV}^2$ and the dash-line for $\kappa_1 = -0.2 ~\mbox{GeV}^2$}
\end{figure}

\begin{figure}
\label{fig4}
\caption{The correlation between $\tau(\Lambda_b)/\tau(b^0)$, $B_{SL}$ and $n_c$. 
Here $x$ stands for $B_{SL}$, $y$ for $n_c$ and $z$ for $\tau(\Lambda_b)/\tau(B^0)$.}
\end{figure}

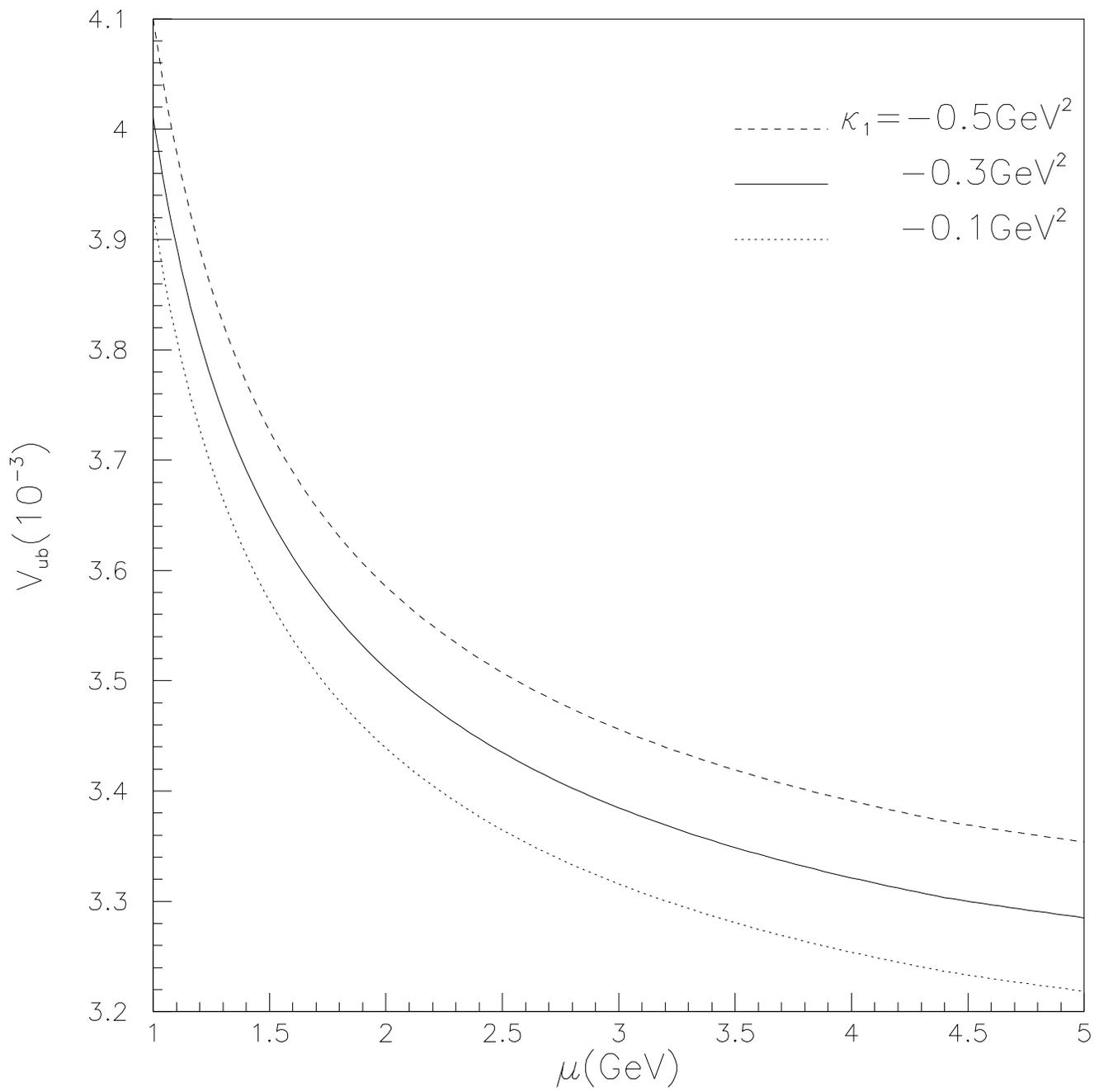
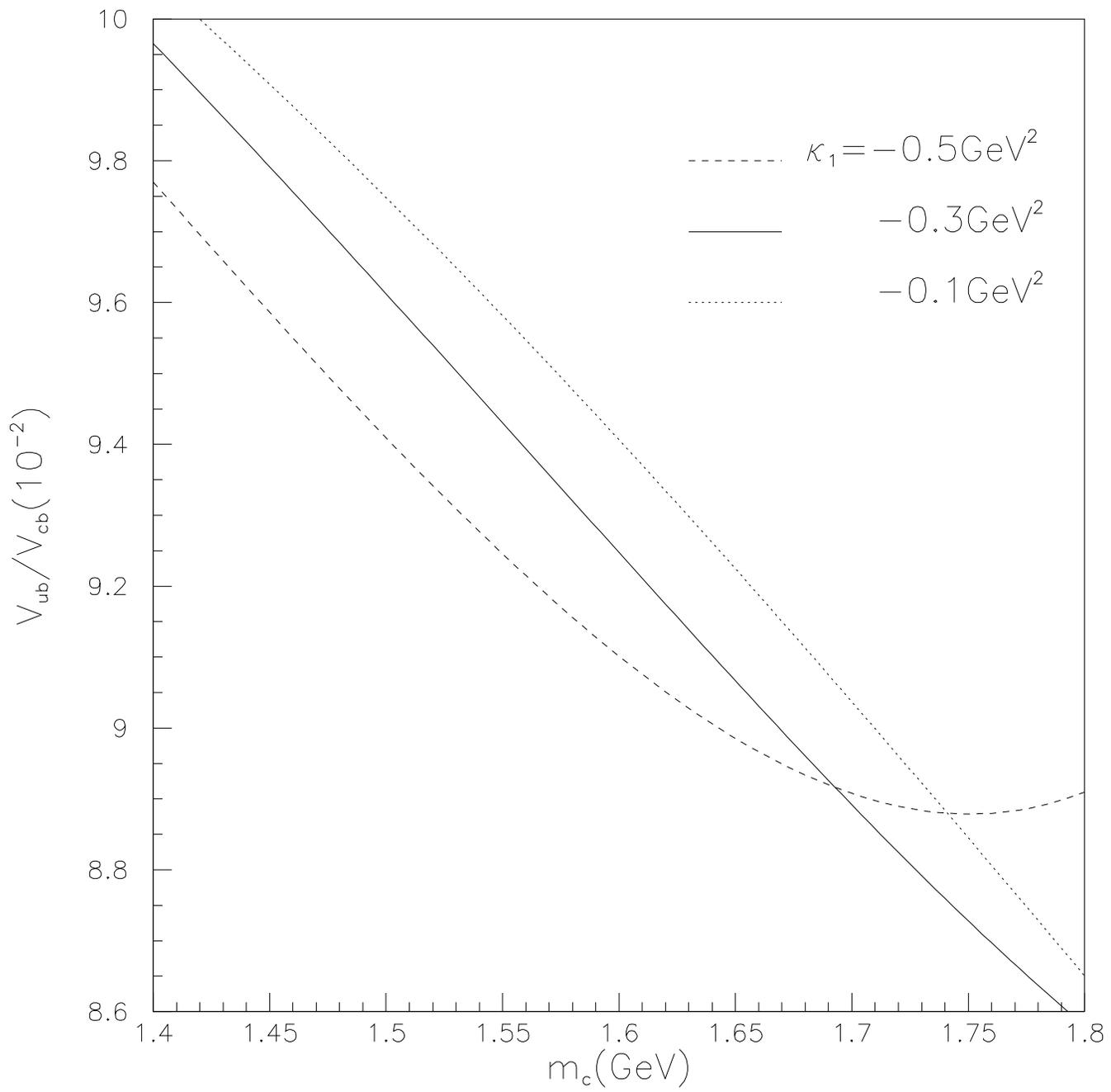
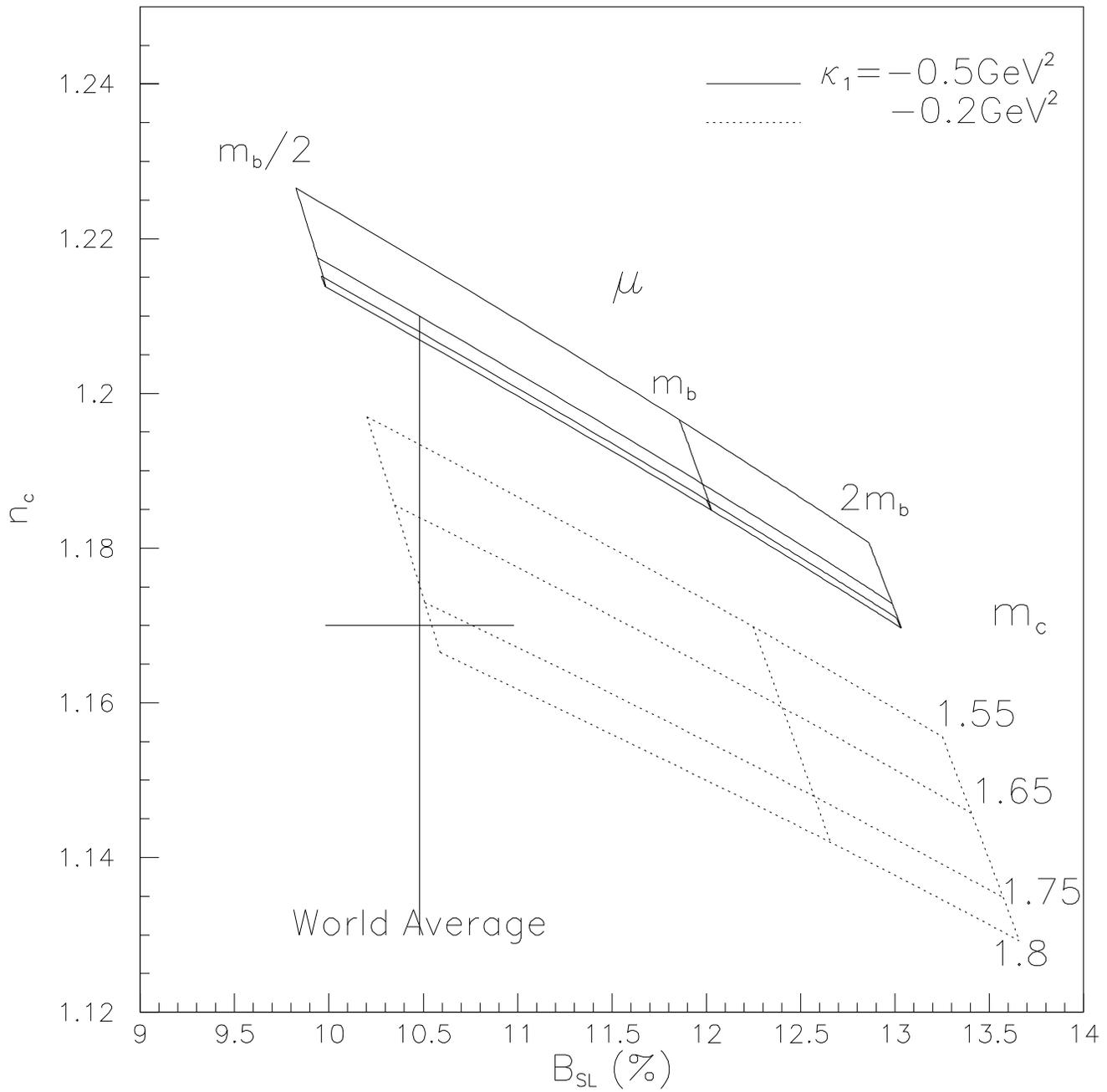
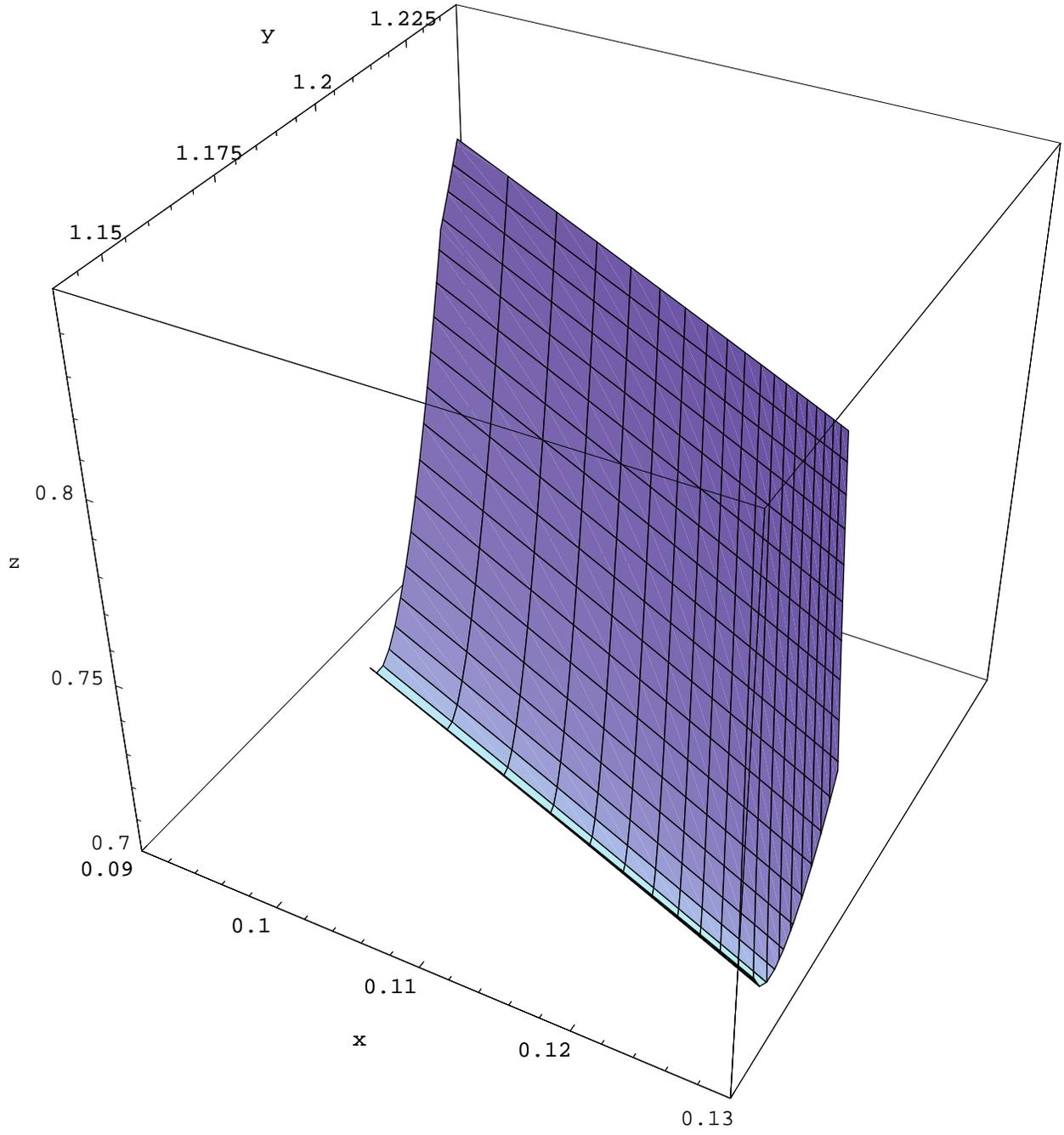
\end{document}